\documentclass[
]{ceurart}

\usepackage{listings}
\usepackage{hyperref} % for autoref
\begin{document}

\copyrightyear{2024}
\copyrightclause{Copyright for this paper by its authors.
  Use permitted under Creative Commons License Attribution 4.0
  International (CC BY 4.0).}

\conference{HCI SI 2023: Human-Computer Interaction Slovenia 2023, January 26, 2024, Maribor, Slovenia}

\title{Teach Me How to ImproVISe: Co-Designing an Augmented Piano Training System for Improvisation}

%\tnotetext[1]{You can use this document as the template for preparing your publication. We recommend using the latest version of the ceurart style.}

\author[1,2]{Jordan Aiko {Deja}}[%
orcid=0000-0001-9341-6088,
email=jordan.deja@famnit.upr.si,
%url=https://www.jrdndj.com/,
]
\cormark[1]
\author[3]{Sandi Štor}[%
%orcid=0000-0002-7784-1356,
%email=sandistor@gmail.com,
%url=https://www.klen.si,
]
\author[4,5]{Ilonka Pucihar}[%
%orcid=0000-0002-7784-1356,
%email=ilonka.pucihar@gmail.com,
%url=https://www.klen.si,
]
\author[1,6,7]{Klen {Čopič Pucihar}}[%
orcid=0000-0002-7784-1356,
email=klen.copic@famnit.upr.si,
%url=https://www.klen.si,
]
%\address[3]{Vrije Universiteit Amsterdam, De Boelelaan 1105, 1081 HV Amsterdam, The Netherlands}
\author[1,7]{Matjaž Kljun}[%
orcid=0000-0002-6988-3046,
email=matjaz.kljun@famnit.upr.si,
%url=https://pim.famnit.upr.si/wp/,
]
%\address[4]{University of Skövde, Högskolevägen 1, 541 28 Skövde, Sweden}

\address[1]{University of Primorska, Faculty of Mathematics, Natural Sciences and Information Technologies, Koper, Slovenia}
\address[2]{De La Salle University, Manila, Philippines}
\address[3]{JazzObala, Portorož, Slovenia}
\address[4]{Vrhnika Music School, Vrhnika, Slovenia}
\address[5]{University of Ljubljana, Academy of Music, Ljubljana, Slovenia}
\address[6]{Faculty of Information Studies, Novo Mesto, Slovenia}
\address[7]{Stellenbosch University, Department of Information Science, Stellenbosch, South Africa}

% %% Footnotes
% \cortext[1]{Corresponding author.}
%\fntext[1]{These authors contributed equally.}

\begin{abstract}
Improvisation is a vital but often neglected aspect of traditional piano teaching. Challenges such as difficulty in assessment and subjectivity have hindered its effective instruction. Technological approaches, including augmentation, aim to enhance piano instruction, but the specific application of digital augmentation for piano improvisation is under-explored. This paper outlines a co-design process developing an Augmented Reality (AR) Piano Improvisation Training System, \textit{ImproVISe}, involving improvisation teachers. The prototype, featuring basic improvisation concepts, was created and refined through expert interaction. Their insights guided the identification of objectives, tools, interaction metaphors, and software features. The findings offer design guidelines and recommendations to address challenges in assessing piano improvisation in a learning context.
\end{abstract}

\begin{keywords}
  Augmented Reality \sep
  Piano \sep
  Improvisation \sep
  Training System \sep
  Music Learning
\end{keywords}

\maketitle

\section{Introduction and Background}

\par Improvisation is an important music skill~\cite{deja2021encouraging} yet tends to be overlooked within conventional piano instruction methods~\cite{deja2022survey}. The challenges of assessing and judging improvisation skills, along with subjective factors, have made it difficult to actually teach musical improvisation especially for novices. For seasoned piano players, demonstrating improvisational skills signifies a broad musical vocabulary. Piano teaching experts argue that improvisation supports creativity of piano learners and performers at all levels. Teaching improvisation to novices and experienced performers enhances rhythmic accuracy, note-reading, concentration, self-reflection, imagination, and bolsters confidence~\cite{chyu2004teaching}.

\par It is known that using digital augmentation (e.g. augmented avatars, rolling visualisations) is effective in teaching musical concepts such as hand-finger-arm synchronisation~\cite{barakonyi2005augmented}, increasing motivation~\cite{rogers2014piano} and even supporting proper sight-reading~\cite{chiang2015oncall}. However, to the best of our knowledge, using augmentation to encourage users to improvise on the piano (as well as other instruments) remains unexplored~\cite{deja2022piano}. To effectively evaluate this potential, it is crucial to understand how to properly-design AR training systems for teaching this skill. Considering there is limited research alone on piano augmentations on improvisation, we posit that involving piano teachers in a co-design process can help in the development of a training system specifically-crafted to support improvisation teaching. We borrow from principles and techniques learned in prior works on co-designing musical instruments and interfaces as seen in ~\cite{turchet2018co, chan2019applying, cook2004remutualizing}. 

\par In summary, this paper presents the following contributions: a) \textbf{narratives from our co-design process of an AR training system for piano improvisation} and its features, which we refer to as \textit{ImproVISe}, and b) \textbf{guidelines and recommendations} to address challenges in assessing piano improvisation in the context of learning.

\section{Co-Desigining ImproVISe}
%\subsection{Initial ImproVIS Features}
\begin{figure}
  \includegraphics[width=\linewidth, height=5.5cm]{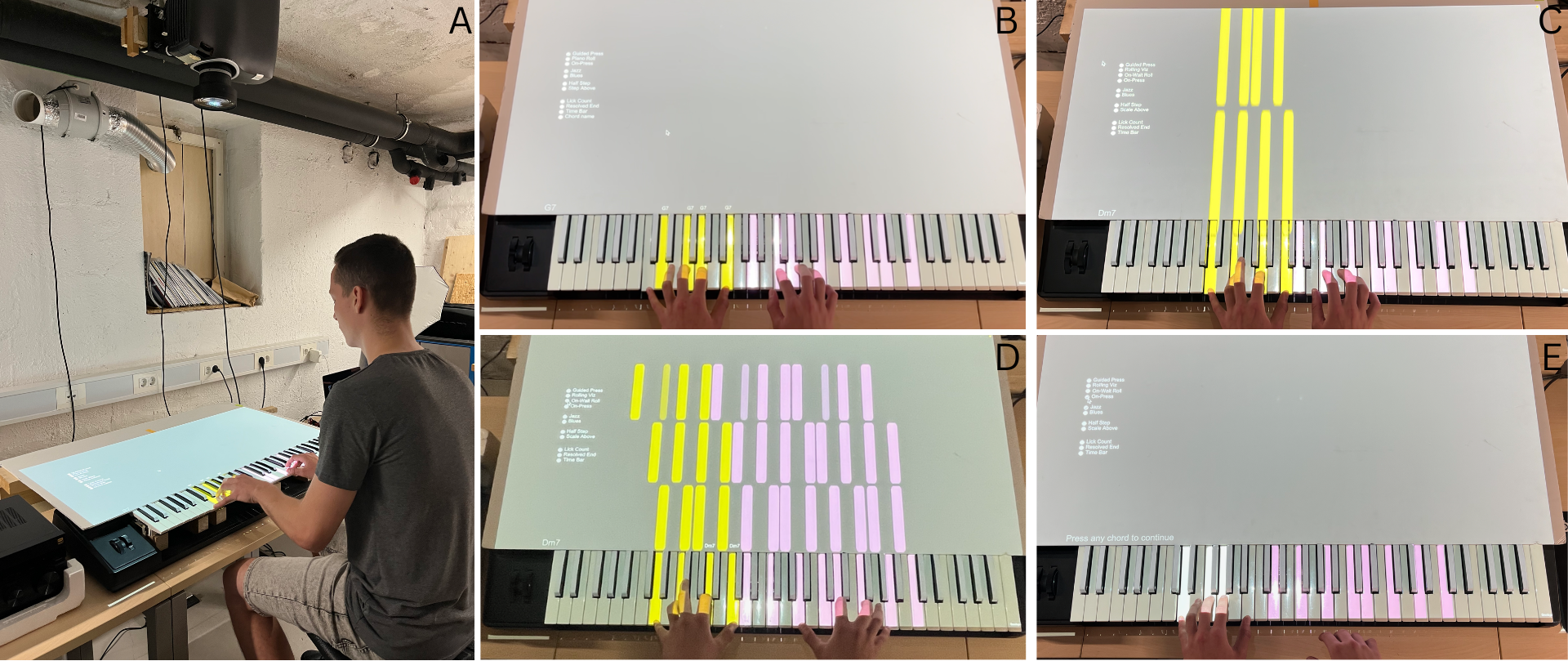}
  \caption{Features of \textit{ImproVISe}: A - Augmented piano space with projector, B - Guided Press, C - Rolling Improv, D - OnWait Improv Rolling, E - Expert Press visualisations.}
  %\Description{A photo of an augmented piano with annotations and labels of the important theme-related features.}
  \label{fig:initprot}
\end{figure}

\par In our co-design, we followed a standard iterative procedure, having distinct phases: a) conceptualising the design, b) translating the design into a high-fidelity prototype, c) engaging with the prototype, and d) refining the prototype based on acquired insights. The first author spearheaded this process with the mentorship of the fourth and fifth authors. The second and third authors are the (impro) experts in the domain of jazz/pop/classical improvisation and teaching improvisation. They have actively participated in all phases of the development of the prototype, offering valuable feedback, and contributing their expertise to the process of the designing the training system.

%lesson concepts
\par For the initial version of \textit{ImproVISe}, we designed and implemented an interactive space inspired by prior works~\cite{rogers2014piano, weing2013piano}. We connected an overhead projector, a modified Clavinova with raised keys and a laptop computer in an interactive space (see \autoref{fig:initprot} A). We built a setup where animated visualisations are projected on the surface, and light certain keys on the Clavinova. These visualisations are programmed in a \texttt{Unity} application specifically designed for this setup. 

\par The different visualisations projected and animations represent the different lessons and metaphors in piano improvisation. Together with the impro experts, we considered basic concepts such as harmony (specifically \texttt{ii-V-i} and \texttt{ii-V-i-VI} progressions), chord tone soloing (also known as licks), and approaches (e.g., half-step and scale-above). These visualisations were implemented in various lessons (which we refer to as modes) in the system namely \texttt{Guided Press}, \texttt{Rolling Improv}, \texttt{OnWait Roll} and \texttt{Express Press} mode. 

% features mapped to concepts
\par The \texttt{Guided Press} mode (\autoref{fig:initprot}B) displays harmonic progressions and chord tones as highlighted keys, allowing users to choose which keys to press~\cite{das2017music}. In yellow, are the keys in the harmonic progression while in pink are the ``musically-correct'' keys that may be pressed for improvisation. In the \texttt{Rolling Improv} mode (\autoref{fig:initprot}C), harmonic progressions appear as falling piano keys in yellow (similar to~\cite{weing2013piano, rogers2014piano}), then revealing chord tones (in pink, like in \texttt{Guided Press} mode) when the right time is triggered. \texttt{OnWait Rolling} mode (\autoref{fig:initprot}D) mirrors \texttt{Rolling Improv} but without the element of timing, allowing the learner to focus on concept absorption without time pressure but with priming~\cite{moro2020performer}. Finally, \texttt{Expert Press} mode (\autoref{fig:initprot}E) assumes user familiarity with chords, lighting up chord tones (in pink) when a harmonic chord is recognised. Approaches (e.g. half-step or scale-above) are implemented as additional guidance across all modes that the user can toggle on or off anytime and appear in a darker shade of purple. 
%\subsection{Co-Design Protocol}

\begin{figure}
  \includegraphics[trim={0 9.5cm 0 0cm}, clip, width=\linewidth]{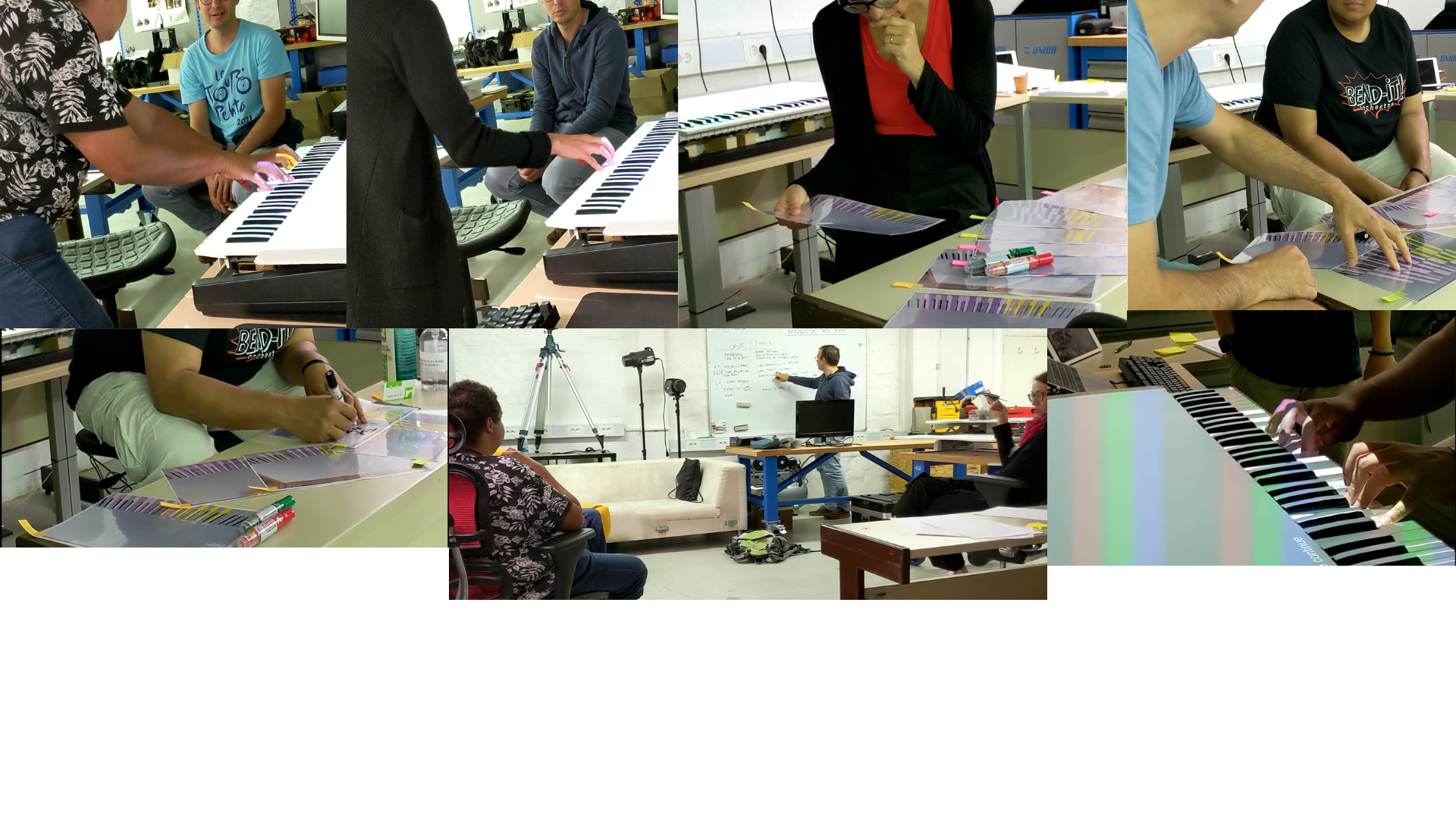}
  \caption{Some impressions during the co-design sessions: interacting with the prototype, planning the lessons, giving feedback, identifying lessons and features. }
  %\Description{A photo of an augmented piano with annotations and labels of the important theme-related features.}
  \label{fig:impressions}
\end{figure}

\par After implementing the first initial version, we then invited the second and third authors to an interactive session (another set of phases c) and d)) with \textit{ImproVISe} (as seen in ~\autoref{fig:impressions}). The co-design sessions consisted of a mix of different steps such as 1) using the prototype, 2) interview on best practices and steps, 3) lesson building and 4) mapping of metaphors. Feedback and insights were collected using multiple modalities (e.g. video, audio, notes, post-it's). After every interactive session, we went again through the phases c) and d) resulting to a total of at least four co-designing sessions. %as seen in ~\autoref{fig:artefacts}). 

% \begin{figure}
%   \includegraphics[width=\linewidth]{figures/artefacts.png}
%   \caption{During co-design sessions, experts engaged with \textit{ImproVIS}. Screenshots of \textit{ImproVIS} features were printed and laminated for annotation. Codes from the sessions, organized into lessons, metrics, and improvisation concepts, influenced the next design of the prototype for an upcoming empirical study.}
%   %\Description{A photo of an augmented piano with annotations and labels of the important theme-related features.}
%   \label{fig:artefacts}
% \end{figure}

\section{Insights Learned and Discussion}

\par The impro experts noted that important musical elements such as timing and patterns were not very obvious in the initial version of \textit{ImproVISe}. While different viz modes are adequate representations of the lessons that they are teaching (e.g. highlights for soloing, piano roll for timing and priming), being able to teach improvisation requires that the proper context is provided to the learner. The experts noted that being confident to improvise is akin to how a non-native speaker becomes confident when speaking a new language. From the repetitive engage-refine sessions with our impro experts, newer lessons and approaches emerged. These were then mapped into tools that can be converted into lessons which in turn can be implemented in the next iteration of the prototype. 

\par From these sessions, \textit{ImproVISe} had to be re-organised into a different structure involving not just lessons but also objectives, tools and new content to generate visualisations with. This in turn will give the learner enough context and practice on how to use them. \autoref{tab:lessons} lists the mapping of these lessons, tools and metaphors. Experts noted that the goal of the system should support rather than replace traditional teaching methods. In addition, understanding the context behind each concept should still align with the prescribed theory and principle of jazz improvisation (e.g. having metronome~\cite{dahlstedt2015mapping}, having musical sheets~\cite{sandnes2019enhanced} etc).  

% \usepackage{graphicx}
\begin{table}[]
\centering
\caption{Overview of Learnings and Features from Co-Design Sessions}
\label{tab:lessons}
\resizebox{\textwidth}{!}{%
\begin{tabular}{|l|l|l|}
\hline
\textbf{Objectives}                                                                      & \textbf{Tools}                                                                                                                                                            & \textbf{Feature Lesson}                                                                           \\ \hline
\begin{tabular}[c]{@{}l@{}}Learning modes\\  and extensions\end{tabular}                 & \begin{tabular}[c]{@{}l@{}}Show different modes,\\ play modes in swing\end{tabular}                                                                                       & Lesson 01: Swing                                                                                  \\ \hline
Understand motifs                                                                        & \begin{tabular}[c]{@{}l@{}}Repeat motifs, sequence the motifs, \\ learn how to form new motifs in Dorian scale.\end{tabular}                                              & Lesson 02: Motifs                                                                                 \\ \hline
\begin{tabular}[c]{@{}l@{}}Be familiar with\\ different rhythmic\\ patterns\end{tabular} & \begin{tabular}[c]{@{}l@{}}Practice with an audio accompaniement,\\ repeat and invent motifs, questions and answers\end{tabular}                                          & Lesson 03: Rhythmic patterns                                                                      \\ \hline
Learn phrases                                                                            & \begin{tabular}[c]{@{}l@{}}Apply and learn a chosen chord progression, \\ learn chord tones, repeat phrases over the chords.\end{tabular}                                 & \begin{tabular}[c]{@{}l@{}}Lesson 04: Relationship between \\ the melody and harmony\end{tabular} \\ \hline
\begin{tabular}[c]{@{}l@{}}Learn basic \\ composition techniques\end{tabular}            & \begin{tabular}[c]{@{}l@{}}Repeat questions and repeat answers, \\ ask question and give your own answer, \\ apply modes and be familiar with the vocabulary\end{tabular} & \begin{tabular}[c]{@{}l@{}}Lesson 05: Composition \\ (Sequence, Q\&A, Variation)\end{tabular}     \\ \hline
Apply different styles                                                                   & \begin{tabular}[c]{@{}l@{}}Apply rhythmic patterns.\\ use tools and all above lessons\end{tabular}                                                                        & \begin{tabular}[c]{@{}l@{}}Lesson 06: Improvise \\ (Compose in the moment)\end{tabular}           \\ \hline
\end{tabular}%
}
\end{table}

\par The iterative co-design sessions also resulted in specific set of practices that were composed and recorded into MIDI files by one of the experts. Every time a new set of files are composed and recorded, they are incorporated into the \textit{ImproVISe}.  

% \begin{figure}
%   \includegraphics[width=\linewidth]{figures/latest.png}
%   \caption{The latest \textit{ImproVIS} version integrates co-design session insights, allowing dynamic content based on teacher-designed motifs and variations.}
%   %\Description{A photo of an augmented piano with annotations and labels of the important theme-related features.}
%   \label{fig:artefacts}
% \end{figure}

\section{Conclusion and Future Work}

\par In this work, we presented the initial features of \textit{ImproVISe}, an AR training system aimed towards encouraging piano learners to improvise. We also narrate the learnings from our co-design sessions which informs the features of the next generation of improvisation training systems. We intend to complete the prototype with the other features identified and perform a longitudinal study to determine whether controlled usage of the prototype can support piano improvisation learners.


\begin{acknowledgments}
We thank Matija Ratković for playing the role of a piano user in the videos and photos used in this paper. 
\end{acknowledgments}

%%
%% Define the bibliography file to be used
\bibliography{main}

\appendix


% The sources for the ceur-art style are available via
% \begin{itemize}
% \item \href{https://github.com/yamadharma/ceurart}{GitHub},
% % \item \href{https://www.overleaf.com/project/5e76702c4acae70001d3bc87}{Overleaf},
% \item
%   \href{https://www.overleaf.com/latex/templates/template-for-submissions-to-ceur-workshop-proceedings-ceur-ws-dot-org/pkfscdkgkhcq}{Overleaf
%     template}.
% \end{itemize}

\end{document}